\documentstyle[aps,prd,epsf]{revtex} 

\draft

\begin{document}
\twocolumn[\hsize\textwidth\columnwidth\hsize\csname
@twocolumnfalse\endcsname

\title{Stochastic optimization methods for extracting cosmological
parameters from CMBR power spectra}

\author{Steen Hannestad}

\address{Institute of Physics and Astronomy,
University of Aarhus,
DK-8000 \AA rhus C, Denmark}

\date{\today}

\maketitle

\begin{abstract}
The reconstruction of the CMBR power spectrum from a map represents
a major computational challenge to which much effort has been 
applied. However, once the power spectrum has been recovered there
still remains the problem of extracting cosmological parameters
from it. Doing this involves optimizing a complicated function in
a many dimensional parameter space. Therefore efficient algorithms
are necessary in order to make this feasible. We have tested several
different types of algorithms and found that the technique known
as simulated annealing is very effective for this purpose.
It is shown that simulated annealing
is able to extract the correct cosmological
parameters from a set of simulated power spectra, but even with
such fast optimization algorithms, a substantial computational
effort is needed.
\end{abstract}

\vskip1.9pc]


\section{Introduction}
In the past few years it has been realized that
the Cosmic Microwave Background Radiation (CMBR) holds information about
virtually all relevant cosmological parameters
\cite{jungman,EHT1}. 
The shape and amplitude
of the fluctuations in the CMBR are strongly dependent on such
parameters as $\Omega$, $H_0$ etc. \cite{tegmark}.
Given a sufficiently accurate map of fluctuations it should therefore
in principle be possible to extract information on the values
of these parameters.
In general, it is customary to describe the fluctuations in spherical
harmonics
\begin{equation}
\frac{\Delta T}{T}(\theta,\phi) =\sum_{lm}a_{lm}Y_{lm}(\theta,\phi),
\end{equation}
where the $a_{lm}$ coefficients are related to the power spectrum by
$C_l = \langle a^*_{lm} a_{lm} \rangle_m$.
For purely Gaussian fluctuations the power spectrum contains all statistical
information about the fluctuations \cite{tegmark}.

The CMBR fluctuations were first detected in 1992 by the COBE satellite
\cite{COBE},
and at present the COBE measurements together with a number of smaller
scale experiments \cite{smaller}
make up our experimental knowledge of the CMBR
power spectrum.
These data are not of sufficient accuracy to really pin down any of the
cosmological parameters, but the next few years will hopefully see
an explosion in the amount of experimental data. Two new satellite
projects, the American MAP and the European PLANCK
\cite{MAPPLANCK}, 
are scheduled and are designed to measure the power spectrum
precisely down to very small scales ($l \simeq 1000$ for MAP and
$l \simeq 2000$ for PLANCK). This should yield sufficient information
to determine almost all relevant cosmological parameters.

However, using CMBR data to extract information about the
underlying cosmological parameters will rely heavily on our ability to
handle very large amounts of data
(Refs.~\cite{tegmark2,bond1,oh,TTH97,tegmark3,GHW98,THBT97} 
and references therein).
The first problem lies in constructing a power spectrum from the
much larger CMBR map. If there are $m$ data points, then the power spectrum
calculation involves inversion of $m \times m$ matrices (an order $m^3$
operation). For the new satellite experiments $m^3$ is prohibitively large
\cite{tegmark2,bond1,oh,TTH97,tegmark3,GHW98,THBT97},
and much effort has been devoted to finding methods for reducing this number
 by exploiting inherent symmetries in the CMBR
\cite{bond1,oh}. 
However, once the power spectrum has been constructed the troubles are
not over. Then the space of cosmological parameters has to be searched
for the best-fit model. If there are $n$ free cosmological parameters,
each sampled by $q$ points, then the computational time scales as
$q^n$ and, if $n$ is large, the problem becomes intractable.
In the present paper we assume that a power spectrum has been constructed,
so that only the problem of searching out the cosmological parameter
space remains.

In general, parameter extraction relies on the fact that for Gaussian
errors it is possible to build a likelihood function from the
set of measurements \cite{THBT97}
\begin{equation}
{\cal L}(\Theta) \propto \exp \left( -\frac{1}{2} x^\dagger 
[C(\Theta)^{-1}] x \right),
\end{equation}
where $\Theta = (\Omega, \Omega_b, H_0, n, \tau, \ldots)$ is a vector
describing the given point in parameter space. $x$ is
a vector containing all the data points. This vector can represent
either the CMBR map, or the reconstructed power spectrum points.
$C(\Theta)$ is the
data covariance matrix.

Assuming that the data points are uncorrelated, so that
the data covariance matrix is diagonal, 
this can be reduced to the simple expression,
${\cal L} \propto e^{-\chi^2/2}$, where
\begin{equation}
\chi^2 = \sum_{l=1}^{N_{\rm max}} \frac{(C_{l, {\rm obs}}-C_{l,{\rm theory}})^2}
{\sigma(C_l)^2},
\label{eq:chi2}
\end{equation} 
is a $\chi^2$-statistics and $N_{\rm max}$ 
is the number of power spectrum data
points \cite{jungman,oh}.

In order to extract parameters from the power spectrum
we need to minimize $\chi^2$ over the multidimensional parameter space.

In general there is no easy way of doing this. The topology
of $\chi^2$ could be very complicated, with several different local
minima. However, let us for now ignore this possible
problem and assume that the
function is unimodal. Then there exist a vast number of algorithms for
extremizing the function. The most efficient methods for optimization
usually depend on the ability to calculate the gradient of the objective
function, $\chi^2$. 
These methods work on completely general continuously
differentiable functions, but under
the right assumptions, $\chi^2$ possesses qualities which makes it
possible to improve on the simple gradient methods.
In general, the second derivative of $\chi^2$ with respect to parameters
$i$ and $j$ is
\begin{equation}
\frac{\partial^2 \chi^2}{\partial \theta_i \partial \theta_j}
 = 2 \sum_{k=2}^N \frac{1}{\sigma_k^2} \left[
\frac{\partial C_l}{\partial \theta_i}
\frac{\partial C_l}{\partial \theta_j} - 
(C_{l, {\rm obs}} - C_l) \frac{\partial^2 C_l}{\partial \theta_i \partial
\theta_j}\right]
\label{eq:second}
\end{equation}
Sufficiently close to the minimum of $\chi^2$, the second term in the
equation above should be small compared with the first. In practice this
means that we get the second derivative information ``for free'' by just
calculating the first derivative. Therefore, if we assume that 
the starting point for the optimization is sufficiently close to the
true minimum, an algorithm utilising second-derivative information
should converge much faster than a gradient method. 
The most popular algorithm of this type
is the Levenberg-Marquardt method \cite{marquardt}.
Note, however, that far away from the minimum, the above expression
for the second derivative can be very wrong and cause the algorithm
to converge much slower.

Both gradient and second order algorithms are typically 
very efficient. However, there are several
weaknesses:
1) They rely on our ability to calculate derivatives of $\chi^2$. 
Although in principle this is no problem, numerical experiments
have shown that results for this derivative are not always reliable
\cite{EHT1}.
For instance, the numerical code for calculating power spectra, CMBFAST
\cite{SZ96},
is fundamentally different for open and flat cosmologies, and has no
implementation of closed models, so that the derivative of $\chi^2$
with respect to $\Omega_0$ is not reliable at $\Omega_0=1$. This is just
one example, but the problem is generic as soon as points are located
sufficiently near parameter boundaries.
2) The next problem is related to the fact that the above methods in 
general works as steepest descent methods. This means that they are very
easily fooled into taking the shortest path towards some local minimum
which needs not be global. If there are either many local minima or the
topology of $\chi^2$ is complicated with many near degeneracies,
then the above gradient-based methods are likely to perform poorly.
Unfortunately this might easily
be the case with any given realization of the CMBR power spectrum.


\section{Stochastic optimization}

\subsection{Multistart algorithms}

The above caveats lead us to look for more robust methods for finding
the true minimum of $\chi^2$. As soon as we are dealing with multimodal
functions it is clear that we cannot contend ourselves with just running
an optimization scheme based on the above method with just one
starting point. The simplest possible improvement on the
above method is a Monte Carlo
multi start algorithm. In this case a starting point is chosen at 
random in the parameter space, and optimization is performed, using
either a gradient or a second-order method. 
After the
algorithm converges a new starting point is chosen. This method
has the advantage that it converges to the global minimum in the
asymptotic limit of infinite computational time.
However, it is easy to improve on it, because the simple multistart
algorithm will detect the same local minimum many times uncritically.

The multi level single linkage (MLSL) algorithm \cite{kan}
tries to alleviate
this problem by mapping out the basins connected with the different 
local minima. If it detects that a trial point lies within
a basin which has already been mapped, then the point is rejected.
Depending on the type of objective function this algorithm can
perform exceedingly well \cite{mlsl}

In what follows we use the simple implementation of the MLSL algorithm
provided by Locatelli \cite{locatelli}. First, we need the following 
definition: Let $x_{\rm max}$ and $x_{\rm min}$ be the maximum and
minimum allowed value of parameter $i$. Then define a new parameter
$q \equiv (x-x_{\rm min})/(x_{\rm max}-x_{\rm min})$, so that 
$q \in [0,1]$. We use this new parameter $q$ in the algorithm below,
so that all cosmological parameters are treated on an equal footing
and the allowed region is a simple hypercube spanning all
values from 0 to 1 in $R^n$. 
The algorithm is then devised
as follow: \\
1) At each step, $k$, pick out $N$ sample points from the allowed
region and calculate the objective function. \\
2) Sort the whole sample of $k N$ points in order of increasing $\chi^2$
value and select the $\gamma k N$ points with smallest values. \\
3) For all of these points, run optimization on given point $q$, iff
\begin{tabbing}
- \= No point $y$ exists so that \\
\> \hspace*{0.5cm}$d(q,y) \leq \alpha$ and $\chi^2(y) \leq 
\chi^2(q)$ \\
- \> $d(q,S) > d$ \\
- \> Optimization was not previously applied to $q$.
\end{tabbing}
Optimization is performed with a gradient method. \\
4) Proceed to step $k+1$.\\

In the above, $d(q,y)$ is the Euclidean distance between $x$ and $y$,
and $S$ is the set of already discovered local minima.
$\alpha$ and $d$ are predefined distances which should be chosen to
optimize the rate of finding local minima.
They are a measure of how large the basins connected with local minima
are in general in that specific problem.
The above method thus includes a host of different parameters which should
be chosen by the user, $N$, $d$, $\gamma$ and $\alpha$. This can make it
quite troublesome to devise an algorithm which performs optimally.
In our implementation we have chosen $N=10$, $\gamma=0.2$, 
$d = 0.1$ and $\alpha=0.1$. Note that this is somewhat in conflict with
the definition given by Refs.~\cite{kan,locatelli}, in that $\alpha$
should really be a quantity which depends on $k$, but in order to 
obtain a simple implementation we have used the above values.

\subsection{Simulated annealing}

A completely different method, which in the next section is shown to 
be very effective for $\chi^2$ minimization on CMBR power spectra,
is simulated annealing.

The method of simulated annealing was
first introduced by Kirkpatrick {\it et al.} in 1983 
\cite{kirk,fleischer}. 
It is based
the cooling behaviour of thermodynamic systems.
Consider a thermodynamic system in contact with a heat bath at
some temperature, $T$. If left for sufficiently long the system 
will approach thermal equilibrium with that temperature.
The heat bath is then cooled, and if this is done slowly
enough the system maintains equilibrium in the cooling phase, and
finally as $T \to 0$ settles into the true ground state, the state
with the lowest possible energy.
This is very similar to global searches for minima
of functions and simulated annealing relies on the fact that the function
to be minimized can be considered as the energy of a thermodynamic 
system. If the system is then cooled from a very high ''temperature''
towards $T=0$ it should find the global minimum, given that it
maintains thermal equilibrium at all times.

In practise one lets the system jump around in parameter space
at random. Given a starting point $i$, a trial point is sought 
according to some prescription, and is then either accepted
or rejected according to the Metropolis acceptance probability
\cite{metropolis}
\begin{equation}
P_{\rm accept}(i+1) = \cases{1 & for $E_{i+1} \leq E_i$ \cr
e^{-(E_{i+1}-E_i)/T} & for $E_{i+1} > E_i$},
\end{equation}
where, in our case $E = \chi^2$.
There are very many similarities between this and thermodynamic systems,
at high temperatures the system visits all states freely, while at low
temperatures it can visit only states very close to the minimum.
For instance it has been shown that by using the above criterion the
system asymptotically approaches the Boltzmann distribution, given that
it is kept at constant temperature asymptotically long \cite{AK89}.
Also, if a system undergoes simulated annealing with
complete thermal equilibrium at all times then as $T \to 0$ the
energy approaches the global minimum \cite{AK89}.
For absolute global convergence to be ensured it is 
thus necessary to allow infinite time
at each temperature.

In order to use simulated annealing for functional optimization it
is necessary to specify three things:\\
1) A space of all possible system configurations \\
2) A cooling schedule for the system \\
3) A neighbourhood structure. \\

Here, the configuration space is a hypercube in $R^n$
bounded by the limits on the individual parameters.

The cooling schedule and the neighbourhood structure are both
something which in general are quite difficult to choose optimally
\cite{fleischer}.
Further, they make the scheme problem dependent. For this reason
adaptive simulated annealing procedures have been devised which
dynamically choose the cooling rate and neighbourhood directly from
the previous iterations in order to maximize the thermalisation rate
\cite{JF95}. 
The problem with this approach is that 
the thermodynamic behaviour is no longer well-defined. For instance
the approach to a Boltzmann distribution is not ensured.

In the present work we choose a relatively simple cooling schedule
and neighbourhood structure, neither of which are adaptive.
In practise we start with an initial temperature, $T_0$, which is then
lowered exponentially by the following criterion
$T_{i+1} = \alpha T_i$,
where $\alpha$ is some constant. 
When the temperature reaches a final value $T_1$ the algorithm stops.
In this way $\alpha$ is a function of the total number of steps, $N_s$,
given as
$\alpha = ({T_1}/{T_0})^{1/N_s}$.

The neighbourhood search is devised
so that at high temperatures the system is prone to make large jumps
whereas at lower temperatures it mostly searches the nearest-neighbour
points. 
In our specific model the parameter space consists of a vector, $\bf x$,
of $n$ free parameters, bounded from below by the vector, ${\bf x}_{\rm min}$,
and from above by ${\bf x}_{\rm max}$.
Let iteration point $i$ have the value $(x_\beta)_i$ for the parameter
labelled $\beta$.
Then the value of this parameter at iteration $i+1$ has 
acceptance probability
given as
\begin{equation}
P[(x_\beta)_{i+1}] \propto
e^{-|(x_\beta)_{i+1}-(x_\beta)_i|/T_{*,\beta}},
\end{equation}
where 
\begin{equation}
T_{*,\beta} = A_\beta [(x_\beta)_{\rm max}-(x_\beta)_{\rm min}](T/T_0)^{1/2},
\end{equation}
and $A_\beta$ is some constant, chosen to yield a good convergence rate. 
The above probability
is set to 0 if $(x_\beta)_{i+1}$ is outside the allowed interval for the given 
parameter.
This criterion for picking out trial points has the desired
quality that it makes large jumps at high temperature and progressively
smaller jumps as the temperature is lowered.
If the objective function depends strongly on $\beta$, then $A_\beta$ should
be small, whereas if it is almost independent of $\beta$, $A_\beta$ should
be large. It is well known that $\chi^2$ is almost degenerate in the
parameter $\Omega_m h^2$ \cite{EHT1}. Therefore it is natural to choose 
$A_{\Omega_m h^2}$ to be small.
In our implementation we have chosen the following values for the
control parameters: $T_0=10^4$, $T_1=2$, $A_{\Omega_m h^2}=1/32$,
$A_\beta=1/8$ for $\beta \neq \Omega_m h^2$.

Note that the method of simulated annealing was first applied to 
simulated CMBR data by Knox \cite{knox}, for a relatively
small model with four free parameters.


\section{Numerical results}

\subsection{Performance of different algorithms}

In order to test the relative efficiency of the different optimization
schemes we have tried to run $\chi^2$ minimization on synthetic
power spectra.
All the power spectra in the present paper have been calculated by use of the
publicly available CMBFAST package \cite{SZ96}.
To make calculations not too cumbersome we have restricted the calculations
to a six-dimensional parameter space, characterised by the vector
$\Theta = (\Omega_m,\Omega_b,H_0,n_S,N_\nu,Q)$.
The model is taken to have flat geometry so that $\Omega_\Lambda = 1-
\Omega_m$. We start from an assumed true model with
$\Theta = (0.5,0.05,50,1,3,30 \, \mu {\rm K})$,
i.e.\ fairly close to the currently favoured $\Lambda$CDM model
\cite{P98}.
Table I shows the free parameters, as well as the allowed region for
each.
We further assume that all $C_l$'s up to $l=1000$ 
can be measured without noise.
That is, the errors are completely dominated by cosmic variance,
with the error being equal to \cite{jungman,tegmark}
\begin{equation}
\sigma(C_l) = \sqrt{\frac{2}{2l+1}}C_l.
\end{equation}
From underlying statistics we have produced a single realisation which
we take to be the measured power spectrum.

Since we have $N=999$ synthetic data points, all normally distributed,
$\chi^2$ of the data set, relative to the true, underlying power
spectrum
should have a $\chi^2$ distribution with mean $N$, and
standard error $\sqrt{2 N}$, so that
\begin{equation}
\chi^2 = 999 \pm 45.
\end{equation}
The specific synthetic data set we use has $\chi^2_* = 1090.98$, i.e.,
it is within about 2$\sigma$ of the expected value.
If the optimization routine is optimal, then for each optimization run
\begin{equation}
\chi^2_{\rm minimization} \leq \chi^2_*.
\end{equation}
\narrowtext
\begin{table}
\caption{The free parameters used in the present analysis, as well as the
allowed range for each.}
\begin{tabular}{lc}
Parameter & Allowed range  \\
\tableline
$Q$ & 5-40 $\mu$K \\
$\Omega_m h^2$ & 0.018-0.49 \\
$\Omega_b h^2$ & 0.002-0.030 \\
$h$ & 0.30-0.75 \\
$n$ & 0.7-1.3 \\
$N_\nu$ & 1-5 \\
\end{tabular}
\end{table}
The average of several optimization runs should preferably yield a value
which is somewhat below $\chi^2_*$.
We therefore have a measure of whether or not the optimization has been
successful.

We have tested four different optimization algorithms on a subset
of the full six-dimensional parameter space. The algorithms are:
Simple Monte Carlo multistart with: 1) gradient optimization method
(G), 2) Levenberg-Marquardt method (LM), 3) multi level single linkage (MLSL),
as described in Section IIa, 4) simulated annealing, as 
described in Section IIb.
Algorithms 1-3 use optimization routines from the PORT3 library
 \cite{gay}.

\begin{figure}
\begin{center}
\epsfysize=10.5truecm\epsfbox{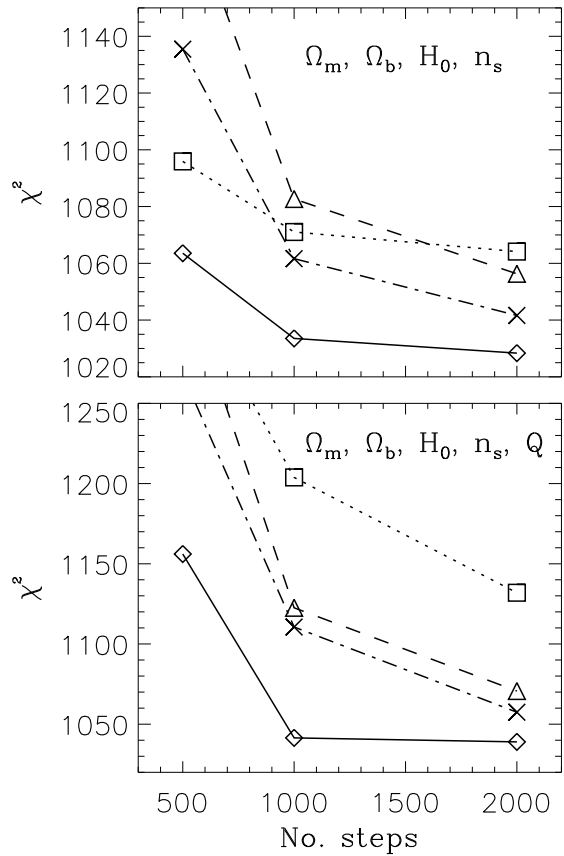}
\vspace*{0.5truecm}
\end{center}
\vspace*{0.5cm}
\caption{The average $\chi^2$ found by the different algorithms.
The data points are for 1) simple gradient method (triangles),
2) Levenberg-Marquardt method (squares), 3) multi level single linkage
(crosses), and simulated annealing (diamonds). The top panel shows
optimization for the case of four free parameters, whereas the bottom
panel shows it for five parameters.}
\label{fig1}
\end{figure}
In order to make direct comparison between the algorithms, we have let them
run for a fixed number of steps, where one step is defined equal to
one power spectrum calculation. All methods, except simulated annealing,
use gradient information, which means that additional power spectra
must be calculated at each iteration. We use two sided derivatives,
so that to calculate the gradient (and Hessian), we need $2 n$ more
calculations, where $n$ is the number of cosmological parameters.
Fig.~1 shows 
the minimum $\chi^2$ found by the different algorithms. Each point
in Fig.~1 stems from a Monte Carlo run of 15 optimizations.

Clearly, the MLSL method improves on the simple multi
start algorithm.
The LM algorithm performs better than gradient optimization in some
cases, but in other cases it is much worse. This is probably
due to
the fact that if the starting point is far away from a local minimum
then the second derivative may yield false information
because Eq.~(\ref{eq:second}) does not hold, 
causing
the algorithm to converge slower.
This weakness could be remedied to some extent
by diagonalising the matrix of 
second-derivatives (Fisher matrix diagonalisation), so that the 
correlation between different parameters is approximately broken.

However, the most striking feature in Fig.~1 is that SA outperforms
the other algorithms easily. Most likely this is due to the fact
that $\chi^2$ possesses valleys where the function has many almost
degenerate local minima. Note that the likelyhood function does not need
to be truly multi-modal for this effect to occur.
It can happen either because the parameter space
is constrained so that the algorithm takes a path which leads out
of the allowed space, or because there are small ``bumps'' on $\chi^2$
close to the global minimum, which cause the gradient algorithms to get
trapped. $\chi^2$ is not multimodal in the sense that it contains
equally good local minima, separated by long distances in parameter
space.
\begin{figure}
\begin{center}
\epsfysize=7truecm\epsfbox{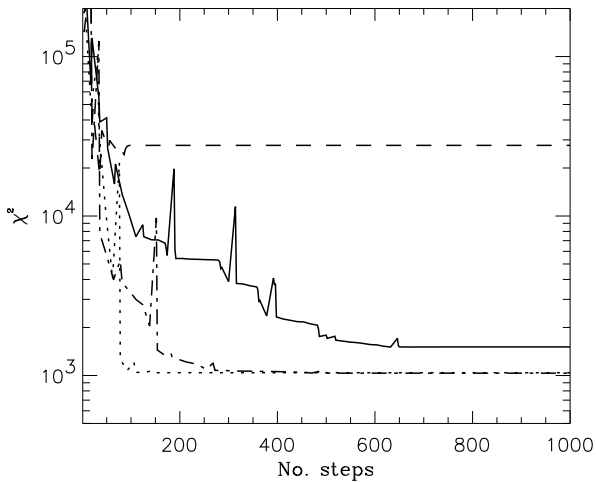}
\end{center}
\baselineskip 17pt
\caption{Four different runs of the simple gradient method (G), without
multi-start.}
\label{fig2}
\end{figure}

For the case of four free parameters (upper panel), most of the
algorithms produce acceptable results with about 1000 steps, 
but with five parameters (lower panel), about 2000 steps are needed.
In both cases, SA needs substantially fewer steps than the other
algorithms.

In Fig.~2 we show four different runs of the simple gradient-based
algorithm without multi-start. In two of the cases the algorithm
converges towards the global minimum, whereas in the two other it 
becomes trapped at much higher lying points in parameter space.
We have tested the effect of varying step size in the gradient calculation
and found that the results do not depend on this.
This figure also shows that the gradient based algorithms generally
converge fairly rapidly (i.e. a few hundred steps), so that the multi-start
algorithm generally runs several times even for relatively a relatively
small number of total steps.

\subsection{Parameter extraction}

If the $\chi^2$ minimization succeeds in finding the global
minimum, then the value found should reflect the underlying measurement
uncertainty. We have performed a detailed Monte Carlo study of
how well the SA algorithm is able to extract parameters from the
power spectrum. The test goes as follows: First, construct $N_{\rm MC}$
synthetic measured power spectra, as described in the previous
section.
Then run optimization on each one of these spectra. This produces
$N_{\rm MC}$ estimated points in parameter space. 
In order to compare these points
with the underlying uncertainty, we then need to calculate
the estimated standard error on the different parameters.
This is done by the standard method of calculating the Fisher information
matrix. At the true point in parameter space, the likelihood function
should be maximal, so that it should have zero gradient. The matrix
of second derivatives is then given by (Eq.~(\ref{eq:second}))
\begin{equation}
I_{ij} = \sum_{l=2}^{l_{\rm max}} (2l+1) C_l^{-2} 
\frac{\partial C_l}{\partial \theta_i}
\frac{\partial C_l}{\partial \theta_j},
\end{equation}
The expected error on the estimation of parameter $i$ is then given by 
\begin{equation}
\sigma_i^2 \simeq (I^{-1})_{ii},
\end{equation}
if we assume that all the relevant cosmological parameters should be
determined simultaneously. 
The expected error on $\Omega_m$ is $\sigma=0.098$, 
given our assumed measurement precision.
Note that above we have again assumed that the only uncertainty
in the measurements is from cosmic variance.

We have performed this Monte Carlo 
test on the 6-dimensional parameter space, using
24 different synthetic spectra. 
We have extracted parameters using SA with a different number
of total steps: 500, 2000 and 4000.
Fig.~3 shows how the estimated points
are distributed for the parameter $\Omega_m$.
We have binned the extracted points in bins of
width 1$\sigma$ up to 5$\sigma$. 
For the optimization performed with 500 steps the distribution
is very wide, showing no specific centering on the true
parameter value. The optimization with 2000 steps
extracts values which are centered on the true
value, indicative of a good optimization. 
Furthermore, the optimization with 4000 steps shows little
improvement over that with 2000, again indicating that the
one with 2000 steps is already performing optimally.
Note that both for 2000 and 4000 steps the distribution
of extracted points is significantly wider than the theoretical
expectation which was calculated assuming a normal distribution
with $\sigma = 0.098$.
One would expect this to be the case since the probability
distribution of any given parameter is only normal close to the
true value, even for a perfect optimization.
Therefore there are likely to be more outlying points
than suggested by the normal distribution.
\begin{figure}
\begin{center}
\epsfysize=10.5truecm\epsfbox{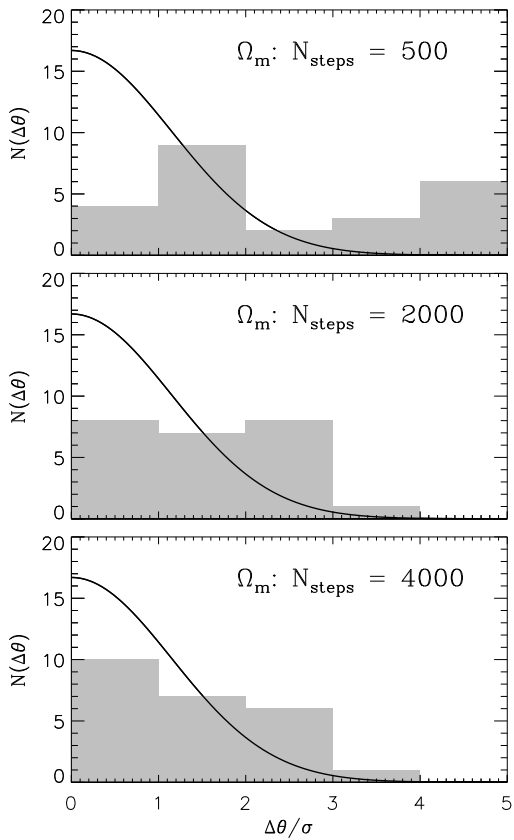}
\vspace*{1.0truecm}
\end{center}
\baselineskip 17pt
\caption{Values of $\Omega_m$ found by the optimization procedure
for the 24 Monte Carlo runs. The recovered values are shown in bins
of width 1$\sigma$, where $\sigma=0.098$ and $\Delta \theta
=| \Omega_{m,{\rm found}}-0.5|$.
The full line is the
theoretical expectation, assuming that errors on $\Omega_m$ are normally
distributed.}
\label{fig3}
\end{figure}

If we have $N_{\rm MC}$ Monte Carlo runs, then if the optimization is perfect
one should obtain a sample mean of roughly
\begin{equation}
\mu_{\rm sample} \simeq \mu_{\rm true} \pm \sigma_s,
\end{equation} 
where $\sigma_s = \sigma/\sqrt{N_{\rm MC}}$
for a given parameter if $N_{\rm MC}$ is large and the 
extracted parameters are drawn from a normal distribution. 
We can also calculate $\chi^2$ for the sample
\begin{equation}
\chi^2_\theta = \sum_{i=1}^{N_{\rm MC}} \frac{1}{\sigma^2}(\theta_{\rm found}-
\theta_{\rm true})^2.
\end{equation}
This function should be approximately
$\chi^2$ distributed. We have calculated
$\mu$ and $\chi^2$ for the sample of extracted parameters, to see if it
is compatible with the theoretical expectations.
Table II shows the values found from the 24 Monte Carlo simulations.
The sample mean found by the optimization with 500 steps deviates by more
than 7$\sigma$ from the expectation. Again this indicates a poor optimization.
The optimizations with 2000 and 4000 steps succeed in recovering
the true mean to within 2$\sigma$.
As for $\chi^2$, it is much lower for the 2000 and 4000 steps optimizations
than for the 500 steps. However, both are still much larger than expected
from a normal distribution. As mentioned above this has to do with the
fact that the distribution is not normal far away from the true
parameter value, so that more outlier points are expected. These contribute
heavily to $\chi^2$, so that a larger value can be expected, even for
a perfect optimization.

As seen above, even for the small 6 parameter model we use, it is necessary
on average to calculate more than 10$^3$ power spectra. Even on
a fast computer this is something which takes several hours.
This must be done each time one wants to check how a new 
proposed cosmological
model fits the data. This very clearly shows the necessity of using
fast optimization algorithms for parameter extraction.

Note that the models we have calculated are flat and without
reionization, including either curvature or reionization
significantly slows the CMBFAST \cite{SZ96} code. Also, more exotic models
like scenarios with decaying neutrinos lead to very cumbersome
CMBR spectrum calculations \cite{kaplinghat}.

The above Monte Carlo method was also used by Knox \cite{knox}
in order
to test the $\chi^2$ optimization efficiency for a small model with 
4 free parameters.
\narrowtext
\begin{table}
\caption{Recovered mean value and $\chi^2$ for the 50 Monte Carlo 
runs performed, for the parameter $\Omega_m$.
Values in parentheses are the expected theoretical values.}
\begin{tabular}{lccc}
 & Steps & $\mu_{\rm sample}$ & $\chi^2$ \\
\tableline
$\Omega_m$ & 500 & 0.655 (0.50 $\pm$ 0.020)  & 198.4 (50 $\pm$ 10) \\
 & 2000 & 0.499 (0.50 $\pm$ 0.020) & 73.7 (24 $\pm$ 6.9) \\
 & 4000 & 0.474 (0.50 $\pm$ 0.020) & 67.7 (24 $\pm$ 6.9) \\
\end{tabular}
\end{table}


\section{Discussion and Conclusions}

We have tested different methods for $\chi^2$ minimization and parameter
extraction on CMBR power spectra. It was found that
simulated annealing
is very effective in this regard, and that it compared very favourably 
with other optimization routines.
The reason for this is most likely that $\chi^2$ posseses very nearly
degenerate minima. Also, numerical noise in the CMBFAST code can cause
the gradient information to become unreliable near stationary points,
causing the gradient based methods to become trapped in points which
are not true minima.

We have also found that even for the simulated annealing algorithm,
many power spectrum calculations are usually necessary in order
to obtain a good estimate of the global minimum. Without a fast
optimization algorithm it is very difficult to extract reliable
parameter estimates from CMBR power spectra, and even with a
routine like SA, it is computationally very demanding as soon as the
parameter space is realistically large (9-10 dimensional).

Note that all of the above calculations rely on stochastic methods
in that they start out at completely random points in the 
allowed parameter space. This is very different from the method used
by Oh, Spergel and Hinshaw \cite{oh}, who use as the initial point a fit
obtained by the chi-by-eye method and then optimize that initial
guess using a second order method. 
This method surely makes the
optimization algorithm converge faster, but suffers greatly from
the problem of how to choose the initial point without biasing
the outcome (i.e. making the algorithm find a minimum which is not 
global). 
We believe that using stochastic optimization is a much more robust way
of optimization.

Interestingly, there are other modern algorithms for optimization
which work along some of the same principles as SA, for instance 
genetic algorithms \cite{genetic}. 
Given the magnitude of the computational challenge
provided by upcoming CMBR data, it appears worthwhile to explore
the potential of such new algorithms.

\acknowledgements
This work was supported by a grant from the Carlsberg Foundation.


\end{document}